\documentstyle[aps,prl,preprint]{revtex}

\title{Speciation as Pattern Formation by
	Competition in a Smooth Fitness Landscape}
\author{Franco Bagnoli\cite{fb}}
\address{Dipartimento di Matematica Applicata, 
Universit\`a di Firenze, via S. Marta,
3 I-50139, Firenze, Italy.} 
\author{Michele Bezzi\cite{mb}}
\address{Dipartimento di Fisica, 
Universit\`a di Bologna, Via Irnerio, 46,
I-40126 Bologna, Italy.}

\begin{document}

\maketitle
\begin{abstract}
We investigate the problem of speciation and coexistence in simple 
ecosystems when the competition among individuals is included 
in the Eigen model for quasi-species.
By suggesting an analogy between 
the competition among strains and the diffusion of a chemical
inhibitor in a reaction-diffusion system,  
the speciation phenomenon
is considered the analogous of chemical pattern formation in genetic space. 
In the limit of vanishing mutation rate 
we obtain analytically the conditions for
speciation. Using different forms of the competition interaction
we show that the
speciation is absent for the genetic 
equivalent of a normal diffusing inhibitor,
and is present for shorter-range interactions.  
The comparison with numerical simulations is very good.
\end{abstract}

In this work we address the problem of speciation (species formation)  in
simple ecosystems, mirroring  some aspects of bacterial and viral evolution.
Our model can be considered as an extension of the Eigen
model~\cite{Eigen71,Eigen:quasispecies}. With respect to the latter, we
introduce  the competition among individuals. 

Eigen's phenomenological theory of self-reproducing macromolecules (or haploid
organisms)
illustrates the concept of stable quasi-species, i.e.\ a peaked distribution
of   genomes around a master sequence,  its width  being determined by
mutations. In its simpler formulation, the various genomes have different
reproductive rates, the logarithm of which constitutes 
the fitness landscape~\cite{Hartle,Wright32,Peliti95}. The master sequence
is located in correspondence of the maximum of the fitness.    
In general  a one to one correspondence between a given phenotype
and a genotype is assumed (no polymorphism nor age structure). 
The genomes are coupled by mutations  
and by a global constraint on the total number
of individuals (constant organization). 
One usually considers only point mutations (the most common ones), which
correspond to a  diffusion process in genetic space.
In this way one can define the concept of distance in genetic space as
the number of mutations needed to connect two genomes.
The Eigen model has also  been studied
in the  contest of statistical mechanics
\cite{Leutheausser,Tarazona,PelitiFranz,Galluccio}.

For a vanishing mutation rate (which is the case for actual organisms), 
only one quasi-species can survive in the long time limit, 
except for the marginal
case of degenerate maxima of the fitness or for an extremely rough landscape
(similar to the spin glass energy landscape) for which the system never
attains equilibrium. Epstein~\cite{Epstein} studied the problem without
considering mutations; he showed that
the coexistence is possible 
if  the species are self-limiting (i.e.\ there exists a
form of  self-competition, modeled for instance by a logistic term) 
and coexisting species does
not compete directly. On the contrary, when two species are in competition 
(because they share some resource --- an enzyme in Epstein's case), only the
fittest one survives. 
However,  he did not introduce the genetic distance among species nor
presented any evolutionary mechanism for the speciation phenomenon.

We think that the direct competition for local resources among strains, 
coupled with a weak mutation rate, 
is the simplest mechanism for modeling both speciation and 
stable coexistence  in 
simple smooth landscapes. The mutations are needed to populate newly formed
niches, while the competition actively separates the strains into
quasi-species.
One can consider the following analogy with a Turing mechanism for chemical
pattern formation. The  main
ingredients are an autocatalytic reaction process (reproduction) with slow
diffusion (mutations) coupled with the emission of a short-lived,
fast-diffusing inhibitor (competition). In this way a local high concentration
of autocatalytic reactants inhibits the growth in its neighborhood, acting as a
local negative interaction. 

In genetic space, the local coupling is given by the competition among
genetically kin individuals. For instance, assuming  a certain distribution
of some resource (such as some essential metabolic component for a
bacterial   population), then the more genetically similar two individuals are,
the wider the fraction of shared resources is. The effects of 
competition on strain $x$ by strain $y$  are modeled by a term proportional to
the relative abundance of the latter, 
$p(y)$, modulated by a function that decreases with the
genetic distance between $x$ and $y$. Another example of this kind of
competition  can be found in  the immune response in mammals. Since the immune
response has a certain degree of specificity, a viral strain $x$ can suffer
from the response triggered by strain $y$ if they are sufficiently near in an
appropriate genetic subspace.  Again, one can think that this effective
competition can be modeled by a term,  
proportional to the relative abundance of the strain that
originated the response, 
which decreases with the genetic distance.

%Our mathematical treatment  can be applied both to chemical and genetic
%problems, and since we work with vanishing mutation rate, it can be applied to
%the modeling of existing ecosystems.  

Although Eigen's model is defined in an hypercubic genetic space, and the error
threshold transition rigorously exists only in an infinite-dimensional space
\cite{Galluccio}, the proposed speciation phenomenon is independent on the
dimension of the genetic space. We shall work therefore in a linear genetic
space.  An instance of a similar (sub-)space in real organisms is  given by a
repeated gene (say a tRNA gene): a fraction of its 
 copies can mutate, linearly varying the
fitness of the individual with the ``chemical composition''
of the gene~\cite{BagnoliLio}. This degenerate case has been
widely studied (see for instance Refs.~\cite{Alves}); one should introduce the
multiplicity of a degenerate state, which can be approximated to a Gaussian, 
but if one works in the neighborhood of 
its maximum (the most common chemical composition) the
multiplicity factors are nearly constants. 
Another example is given by the level of catalytic
activity of a protein.  A linear space has also been used for modeling the
evolution of RNA viruses on HeLa cultures~\cite{Kessler}. 

Let us start with a one dimensional ``chemical'' model of cells that reproduce
asexually  and slowly diffuse (in real space), $p=p(x,t)$ being their relative
abundance at position $x$ and at time $t$. These cells constitutively emit a
short-lived, fast-diffusing mitosys inhibitor $q=q(x,t)$. This inhibitor
may be simply identified with some waste or with 
the consumption of a local resource (say
oxygen).  
The diffusion of the inhibitor is modeled as
\begin{equation}
	\frac{\partial q}{\partial t} = k_0 p +
		 D \frac{\partial ^2q}{\partial x^2} -k_1 q,\label{q}
\end{equation}
where $k_0$, $k_1$ and $D$ are the production, annihilation
and diffusion rates of $q$.

The evolution of the distribution $p$ is given by
\begin{equation} 	
	\frac{\partial p}{\partial t} = \left(A(x, t) -\overline{A}(t)\right)p + 
		\mu \frac{\partial ^2p}
		{\partial x^2},\label{p} 
\end{equation}
\begin{equation} 
	\overline{A}(t) = \int A(y,t)p(y,t){\rm d} y.\label{A}
\end{equation}
The growth rate $A$ can be expressed in terms of the fitness $H$ as
\begin{equation} 
	A(x,t) = \exp\left(H(x,t)\right).\label{H}
\end{equation}
Due to the form of equation~(\ref{p}), the distribution $p$ 
is always normalized to one. The
diffusion rate of $q$, $D$, is assumed to be much larger than $\mu$. The 
growth rate $A$,  can be decomposed in two
factors, $A(x,t)  = A_0(x) A_1(q(x,t))$,
where $A_0$ gives the reproductive rate in absence of $q$, so $A_1(0) = 1$. In
presence of a large concentration of the inhibitor $q$ the reproduction stops,
so $A_1(\infty)=0$. A possible choice is 
\[ 
A(x,t) = \exp(H_0(x) -  q(x,t)).
\] 
For instance, $H_0(x)$ could model the sources of food or, for algae
culture, the distribution of light. 

Since we assumed a strong separation in time scales, we look for a
stationary distribution $\tilde q(x,t)$ of the inhibitor
(eq.~(\ref{q})) by keeping $p$ fixed. This  is given 
 by a convolution of the distribution $p$:
\[
	\tilde q (x,t) = J\int  \exp\left(-\frac{|x-y|}{R}\right) p(y,t) {\rm d}y, 
\] 
where $J$ and $R$ depend on the parameters $k_0$, $k_1$, $D$. In the following
we shall use $J$ and $R$ as control parameters, disregarding their origin.

We can generalize this scenario to non-linear diffusion processes of the
inhibitor by using 
the reaction-diffusion equation eq.~(\ref{p}), with the fitness $H$ 
and the kernel $K$ are given by
\begin{equation}
	H(x,t) = 	H_0(x) - J\int K\left(
			\frac{x-y}{R}\right) p(y,t) {\rm d}y \label{kernel}
\end{equation}
\begin{equation}
	K(r) = \exp\left(-\frac{|r|^\alpha}{\alpha}\right),
	\label{K}
\end{equation}
i.e. a symmetric decreasing function of $r$ with 
$K(0)=1$. The parameters $J$ and $\alpha$ control
the intensity of the competition and 
the steepness of the interaction, respectively.    

Let us consider the correspondence with the genetic space:  the quantity $x$ now
identifies a genome, the diffusion rate $\mu$ is given by mutations, and the
inhibitor $q$ (which is no more a real substance)
represents the competition among genetically related strains. The
effects of competition are much faster than the genetic drift  (mutations), so
that the previous hypotheses are valid. While the genetic  interaction kernel
$K(r)$ is not given by a diffusion process, its general form should be similar
to that of eq.~(\ref{K}): a decreasing  function of the genetic distance between
two strains.  
We shall refer to 
the $p$-independent contribution to the fitness, $H_0(x)$,
as the static fitness landscape.

Our model is thus defined by eqs.~(\ref{p}--\ref{K}). 
We are interested in its asymptotic behavior in the limit
$\mu\rightarrow 0$. Actually, the mutation mechanism is needed only to define
the genetic distance and to allow population of an eventual niche.
The results should not change qualitatively if one includes more realistic
mutation mechanisms.

Let us first examine the behavior of eq.~(\ref{p}) in absence of
competition ($J=0$) for a smooth static landscape and 
 a vanishing mutation rate. This
corresponds to the Eigen model in one dimension: 
since it does not exhibit any phase transition, the
asymptotic distribution is unique. 
The asymptotic distribution is given by one delta function peaked
around the global maximum of the static landscape, or more delta functions 
(coexistence) if
the global maxima are degenerate. 
The effect of a small mutation rate is simply that of broadening
the distribution from a delta peak to a bell-shaped curve~\cite{ECAL}. 

%\begin{figure}[t]
%\psfig{figure=fig1.ps,width=9cm}
%\caption{\small 
%Static fitness $H_0$, effective fitness 
%$H$,  and asymptotic distribution $p$ 
%numerically computed for the following values of
%parameters: $\alpha=2$, $\mu=0.01$, $H_0=1.0$, 
%$b=0.04$, $J=0.6$, $R=10$, $r=3$ and $N=100$.}
%\end{figure}

While the degeneracy of maxima of the  static fitness
landscape is a very particular condition, 
we shall show in the following that in
presence of competition this is a generic case. 
For illustration, we report 
in Figure~1 the numerical computation of the asymptotic behavior of the 
model
for a possible evolutive
scenario that leads to the coexistence of three species. We have chosen a
smooth static fitness $H_0$ (see eq.~(\ref{H0}))
and a Gaussian ($\alpha=2$) competition 
kernel. The effective fitness  $H$ is almost 
degenerate (here $\mu>0$ and the competition effect extends on
the neighborhood of
the maxima), and this leads to the coexistence. One could show that 
the curvature of the maxima affects the width and the height of the
quasi-species distribution in presence of mutations~\cite{ECAL}. 

We shall now derive the conditions for the coexistence of multiple species. 
Let us assume that the asymptotic distribution is formed by $L$ delta
peaks $p_k$, $k=0, \dots, L-1$,
 for a vanishing mutation rate (or $L$ non-overlapping bell shaped
curves for a small mutation rate) centered at $y_k$.
The weight of each quasi species is $\gamma_k$, i.e.
\[
	 \int p_k(x) dx = \gamma_k, \qquad\sum_{k=0}^{L-1} \gamma_k = 1.
\]
The quasi-species are ordered such as  $\gamma_0 \ge
\gamma_1,  \dots, \ge \gamma_{L-1}$. 

 The evolution equations for the $p_k$ are 
($\mu \rightarrow 0$) 
\[
	\frac{\partial p_k}{\partial t} = (A(y_k) - \overline A) p_k,
\]
where $A(x) = \exp\left(H(x)\right)$ and
\[
	H(x) = 
		H_0(x) - J\sum_{j=0}^{L-1} K\left(\frac{x - y_j}{R}\right) \gamma_j.
\]

The stability condition of the asymptotic distribution is 
$(A(y_k) - \overline A) p_k = 0$, i.e. either
 $A(y_k) = \overline A = \text{const}$
(degeneracy of maxima) or $p_k=0$ (all other points). In other terms one can
say that in a stable environment the fitness of all individuals is the same,
independently on the species. 

The position $y_k$ and the weight $\gamma_k$ of the quasi-species
are given by $A(y_k) = \overline A = \text{const}$ and 
$\left.{\partial A(x)}/{\partial x}\right|_{y_k} = 0$, or, in terms of the
fitness $H$, by
\[
	H_0(y_k) - J \sum_{j=0}^{L-1} K\left(\frac{y_k-y_j}{R}\right)
		 \gamma_j = \text{const}
\]
\[
	H'_0(y_k)  - \frac{J}{R}\sum_{j=0}^{L-1} K'\left(\frac{y_k-y_j}{R}\right)
	 \gamma_j = 0
\]

Let us compute the phase boundary for coexistence of three species for two
kinds of kernels: the exponential (diffusion) one ($\alpha=1$)
and a Gaussian one ($\alpha=2$). 

We assume that the static fitness $H_0(x)$ is a symmetric
linear decreasing function
except in the vicinity of $x=0$, where it has a quadratic maximum:
\begin{equation}
	H_0(x) = b\left(
		1-\frac{|x|}{r} - \frac{1}{1+|x|/r}
	\right)\label{H0}
\end{equation}
so that close to $x=0$ one has 
$H_0(x) \simeq -b x^2/r^2$ and for $x\rightarrow \infty$,
$H_0(x) \simeq b(1-|x|/r)$. We have checked numerically that the results are
qualitatively independent on the exact form of the static fitness, providing
that it is a smooth decreasing function. 

Due to the symmetries of the problem, we have one quasi-species at $x=0$ and
two symmetric quasi-species at $x=\pm y$. Neglecting the mutual influence of
the two marginal quasi-species, and considering that $H'_0(0) = K'(0)=0$, 
$K'(y/R) = -K'(-y/r)$, $K(0)=J$ 
and that the three-species threshold is given by $\gamma_0=1$ and $\gamma_1=0$,
 we have 
\[
	\tilde{b}\left(1-\frac{\tilde{y}}{\tilde{r}}\right) 
				- K(\tilde{y}) = -1,  
\]
\[
	\frac{\tilde{b}}{\tilde{r}} + K'(\tilde{y}) = 0.
\]
where $\tilde{y}=y/R$, $\tilde{r} = r/R$ and $\tilde{b} = b/J$. 
In the following we drop the tildes for convenience.
Thus
\[
 r - z - G \exp\left(-\frac{z^\alpha}{\alpha}\right) = -G,
\]
\[
 G z^{\alpha-1}\exp\left(-\frac{z^\alpha}{\alpha}\right) = 1,
\]
where $G=r/b$.

For $\alpha=1$ we have the coexistence condition
\[
 \ln(G) = r -1 + G.
\]
The only parameters that satisfy these equations are $G=1$ and $r=0$,
i.e.\ a
 flat landscape ($b=0$) with infinite range interaction ($R=\infty$). 
Since the coexistence region reduces to a single point,
it is suggested that $\alpha=1$ is a marginal case. 

For $\alpha=2$ the coexistence condition is given by
\[
	z^2 -(G+r)z + 1 = 0,
\]
\[
	Gz\exp\left(-\frac{z^2}{2}\right) = 1.
\]
One can solve numerically this system and obtain the boundary 
$G_c(r)$ for the coexistence. In the limit $r \rightarrow 0$ (static fitness
almost flat) one has 
\begin{equation}
	G_c(r) \simeq G_c(0) - r \label{Gc}
\end{equation}
with $G_c(0) = 2.216\dots$. 

%\begin{figure}[t]
%\psfig{figure=fig2.ps,width=9cm,angle=270}
%\caption{\small Three-species coexistence boundary 
%$G_c$ for $\alpha=2$. The continuous 
%line represents the analytical 
%approximation, eq.~(\ref{Gc}), the circles are obtained from
%numerical simulations. The error bars represent the maximum error.}
%\end{figure}

We have performed several numerical simulations for 
different values of the parameters, whose results are presented in Figure~2.
The boundary of the multi-species phase is well approximated by eq.~(\ref{Gc});
in particular, this boundary does not depends on the mutation rate
$\mu$, at least for $\mu < 0.1$, which can be considered
a very high mutation rate for
real organisms. The most important effect of $\mu$ is the broadening of
quasi-species curves, which can eventually merge.

In conclusion, we have introduced a model for the genetic evolution
of haploid organisms under the pressure of a static fitness landscape 
 and competition. 
This model exhibits the phenomenon of species formation in a way reminiscent of
a chemical pattern formation via a Turing-like mechanism.
We have analyzed analytically this system in the limit of
vanishing mutation rate and linear genetic space,   
showing that an increasing level of a short-range  
competition induces 
a transition from a single species distribution to a stable
environment  in which multiple genetically distinct species are present.
The comparison of the analytical approximation 
with the numerical integration of the original 
differential equations is very good.
We think that the mechanism that we proposed is the simplest one for modeling 
speciation and species coexistence in a smooth (or flat) fitness landscape.

We wish to thank G. Guasti, G. Cocho, L. Peliti, G. Martinez-Mekler and
P.Li\'o for fruitful discussions.
M.B. thanks the Dipartimento di Matematica Applicata ``G. Sansone'' for 
friendly hospitality. Part of this work was done during the workshop
{\it Chaos and Complexity} at ISI-Villa Gualino (Torino, Italy)
under CE contract ERBCHBGCT930295.

\newpage
\section*{Figure captions}

\begin{description}

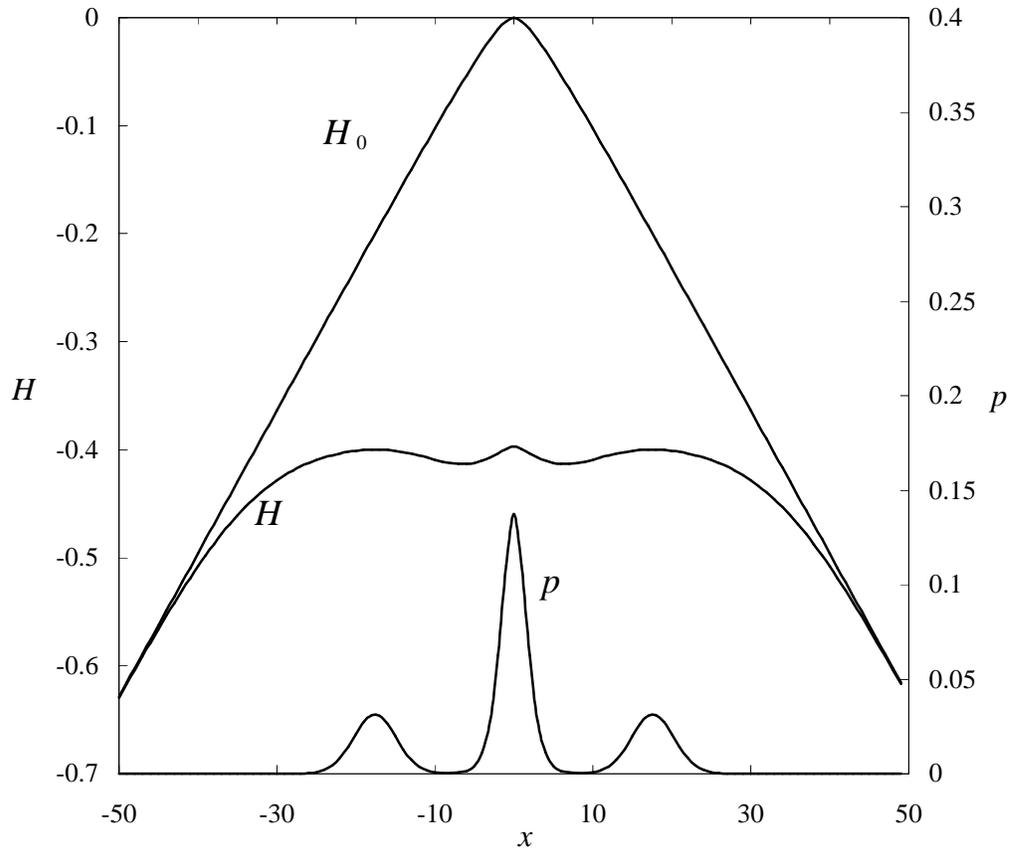
\item{Fig. 1.} 
\label{Potential}
Static fitness $H_0$, effective fitness 
$H$,  and asymptotic distribution $p$ 
numerically computed for the following values of
parameters: $\alpha=2$, $\mu=0.01$, $H_0=1.0$, 
$b=0.04$, $J=0.6$, $R=10$, $r=3$ and $N=100$.

\item{Fig. 2.} 
\label{figG}
Three-species coexistence boundary $G_c$ for $\alpha=2$. The continuous 
line represents the analytical 
approximation, eq.~(\ref{Gc}), the circles are obtained from
numerical simulations. The error bars represent the maximum error.
 
\end{description}

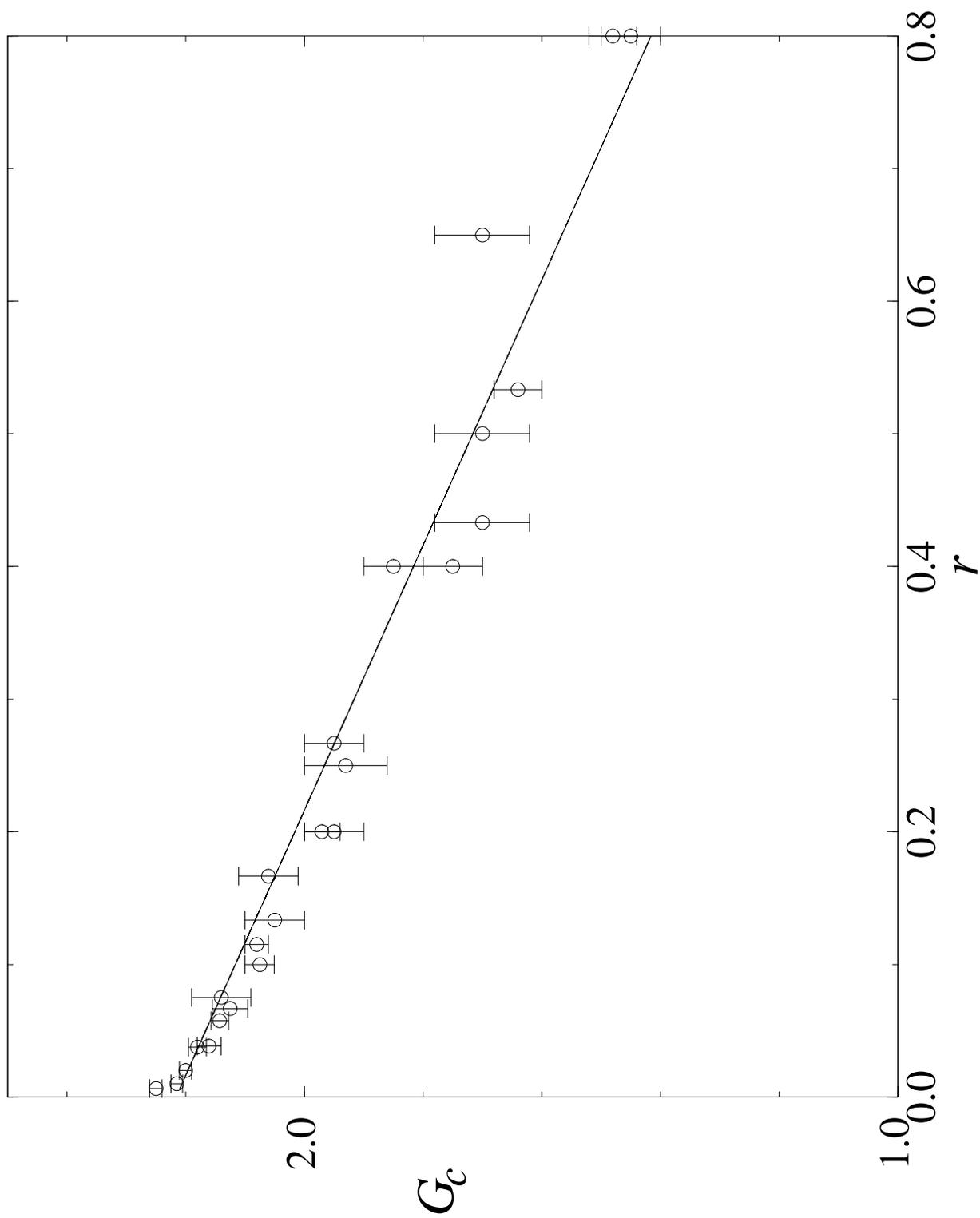
\end{document}